\documentclass[epj]{svjour}
\usepackage{graphics}
\usepackage[utf8]{inputenc}
\usepackage{xcolor}
\usepackage{breakcites}
\usepackage{amsmath}
\usepackage{rotating}
\usepackage{cite}

\newcommand{\Msun}{\rm M_{\rm Sun}}
\begin{document}
\title{Delta Baryons in Neutron-Star Matter under Strong Magnetic Fields}
\author{{Veronica Dexheimer}\inst{1},
{Kauan D. Marquez}\inst{2} and  {Débora P. Menezes\inst{2}}
}                     
%
%
\institute{{Department of Physics, Kent State University, Kent, OH 44243 USA} \and
{Depto de F\'{\i}sica - CFM - Universidade Federal de Santa Catarina  Florian\'opolis - SC - CP. 476 - CEP 88.040 - 900 - Brazil}  }
\date{Received: date / Revised version: date}
%
\abstract{
In this work, we study magnetic field effects on neutron star matter containing the baryon octet and additional heavier spin 3/2 baryons
(the $\Delta$'s). We make use of two different relativistic hadronic models that contain an additional vector-isovector self interaction for the mesons: one version of a relativistic mean field (RMF) model and the Chiral Mean Field (CMF) model. We find that both the additional interaction and a strong magnetic field enhance the $\Delta$ baryon population in dense matter, while decreasing the relative density of hyperons. At the same time that the  
 vector-isovector meson interaction 
modifies neutron-star masses very little ($<0.1~\Msun$), it decreases their radii considerably, allowing both models to be in better agreement with observations. Together, these features indicate that magnetic neutron stars are likely to contain $\Delta$ baryons in their interior.
\PACS{
      {PACS-key}{discribing text of that key}   \and
      {PACS-key}{discribing text of that key}
     } 
} 
\authorrunning{Dexheimer et al.}
\titlerunning{Delta Baryons in Neutron-Star Matter under Strong Magnetic Fields}

\maketitle

\section{Introduction}

The role of the baryon decuplet has been studied in neutron stars in several works, since first investigated by Glendenning in 1982 \cite{Glendenning:1982nc}. More recently, it has appeared in several publications \cite{Oliveira:2019xni,Malfatti:2020onm,Sahoo:2020ipd,Motta:2019ywl,Li:2020ias,Raduta:2020fdn,Ribes:2019kno,Sun:2018tmw}. The energies involved in the core of neutron stars is more than sufficient to create these heavier baryons, the lightest one, $\Delta$, being only $292$ MeV heavier than a neutron (see Fig.~1 in Ref.~\cite{Malfatti:2020onm}). Considering charge neutrality, negatively charged spin $3/2$ baryons are favored, while the positively charged ones are suppressed, in the same way as the hyperons. In this way, the only mechanism that could prevent $\Delta$'s from appearing in neutron stars would be a very repulsive coupling. The interested reader can see Ref.~\cite{Raduta:2021xiz} for a discussion on the stability of $\Delta$-rich matter. Currently, there is very little known about how these particles couple in dense matter, including some potentials extracted from heavy-ion collisions and scattering experiments, resulting in a range for the $\Delta$ potential for symmetric matter at saturation {{(with respect to the nucleon potential)}} of $ U_N - 30$ MeV $\to\ 2/3\ U_N$. See discussions in Refs.~\cite{Drago:2014oja,Kolomeitsev:2016ptu} and references therein for more details. {{Note that recent transport model calculations suggest, on the other hand, that the $\Delta$ potential is different from that of the nucleon \cite{Cozma:2021tfu}.}}

The fact that {{$\Delta$-baryons}} have spin $3/2$ immediately raises the question about how they are affected by the presence of strong external magnetic fields. This was first discussed in Ref.~\cite{dePaoli:2012eq}, but only investigated in the context of neutron stars recently in Ref.~\cite{Thapa:2020ohp}, where the authors used the so-called universal magnetic field profile \cite{Chatterjee:2018prm} in the Tolman-Oppenheimer-Volkoff equations. They found that the magnetic field affects the particle population, enhancing strangeness in neutron stars. 

In the present work, we introduce for the first time a non-trivial (in the sense that is not yet commonly used in the literature) self vector-isovector interaction to describe matter found in the core of neutron stars under the influence of strong magnetic fields. This interaction was shown to improve the agreement of  mean-field models with neutron-skin data \cite{Horowitz:2002mb}, neutron-star radii \cite {Schramm:2002xa,Bizarro:2015wxa,Pais:2016xiu}, symmetry energy slope \cite{Dexheimer:2015qha,Hornick:2018kfi}, tidal deformability, and low-density constraints from chiral effective field theory \cite{Dexheimer:2018dhb}.

In order to generalize our results, we repeat our analysis for two different relativistic hadronic models. The first one is a modified version of the Walecka model within a relativistic mean field (RMF) approximation and the second one, the CMF model, includes chiral symmetry restoration in the expected regime of high energy.

\section{Basic formalism and Results}

\subsection{Magnetic Field}

In order to study modifications introduced by an external magnetic field $B$ on fermions, we modify the calculation of thermodynamical quantities of each {{hadronic and leptonic}} particle species with non-zero electric charge $q$ according to the general procedure of taking
\begin{eqnarray}
\sum_{\rm{spin}}\int d^3k \to \frac{|q| B}{(2\pi)^2} \sum_{\rm{spin}} \sum_{n} \int dk_z ,
\end{eqnarray}
where $k$ is the momentum, $z$ is the local direction of the magnetic field and $n$ is the discretized orbital angular momentum that the charged particle acquires in the plane transverse to $B$. The total effective energy of a charged particle (modified by the repulsive interactions) with effective mass $m^*$ becomes
\begin{eqnarray}
E^*=\sqrt{k_z^2+{m^*}^2+2\nu|q|B} .
\end{eqnarray}
The sum in $n$ at zero temperature goes until a maximum (integer) corresponding to Landau level $\nu$ for which {{${k_z}^2\ge0$}}, i.e.,
\begin{eqnarray}
\nu \le \nu_{\rm max} = \left\lfloor \frac{{E^*}^2 - {m^*}^2}{2|q|B} \right\rfloor ,
\end{eqnarray}
where $\nu=n +\frac{1}{2}-\frac{s}{2}\frac{q}{|q|}$ depends on spin and electric charge. For each of the models described in the following, the particle effective energies and masses include different vector and scalar interactions, {{except for the leptons (electrons and muons), which are treated as a free Fermi gas}}. See Ref.~\cite{Strickland:2012vu} for a complete list of thermodynamics quantities for magnetized fermions at finite and zero temperature, together with a literature review. In what follows, we consider that the strength of the magnetic field $B$ is a fixed quantity in the equation of state (EoS). 
We refer to Refs. \cite{PhysRevC89,Prakash}  for a more detailed discussion of the formalism involved in the description of strong magnetic field effects on the equation of state of models with interactions.

\begin{figure*}
\centering
\includegraphics[angle=270,width=8.9cm]{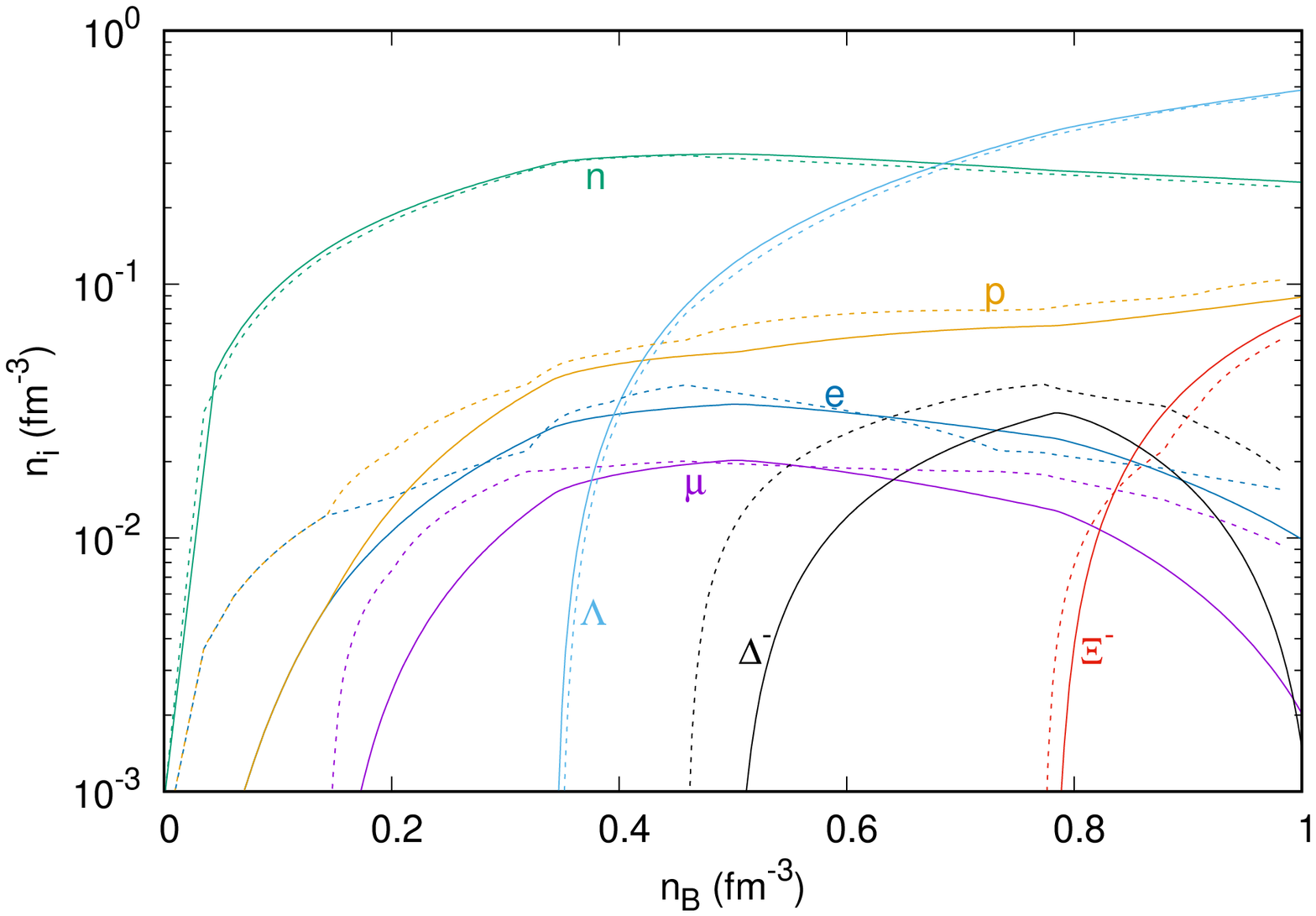}    
\includegraphics[angle=270,width=8.9cm]{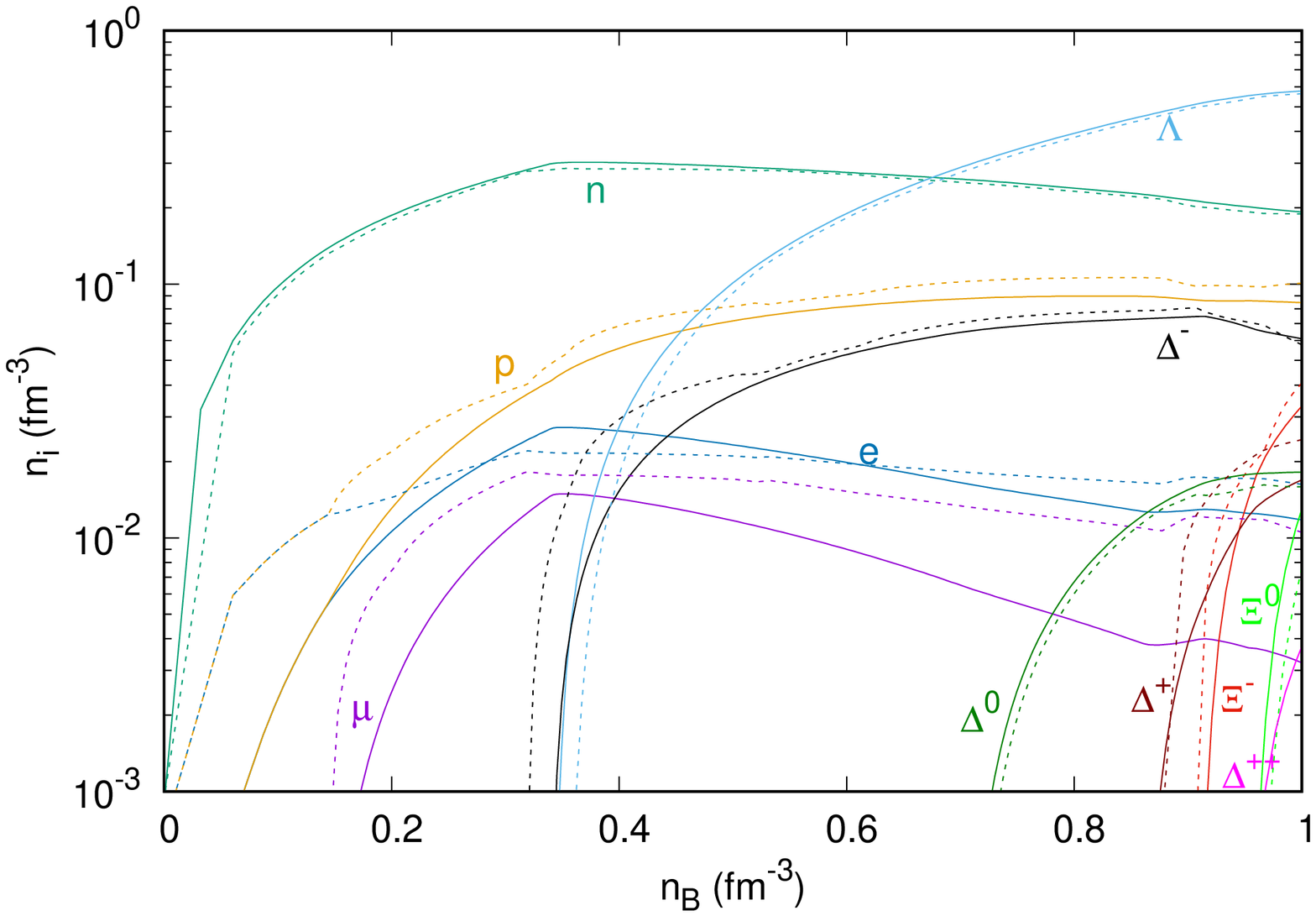}
\caption{Particle population for the RMF model with GM1, $\beta=1.0$ parametrization (left panel) and $\beta=1.1$ parametrization (right panel) as a function of the baryon number density. Full lines show results without magnetic fields, while dashed lines show results including a magnetic field strength of $B=3\times10^{18}$ G.}
\label{popnlwm}
\end{figure*}

The number of Landau levels occupied by particles increases with density and temperature, but decreases with the magnetic field strength, as seen in Ref.~\cite{PhysRevC83}. In the cases we study in this work, the number of Landau levels goes from zero or $1$ at low densities to a number $\leq 5$ (for the proton) at large densities, corresponding to the center of the neutron stars.

\subsection{Relativistic Mean-Field Model}

The relativistic mean-field (RMF) model we discuss is a rather generalized version of the $\sigma-\omega$ hadrondynamics model, in which the strong interaction is emulated by the exchange of scalar-isoscalar meson $\sigma$, the vector-isoscalar meson $\omega$, and the vector-isovector meson $\rho$. In this work, we consider the GM1 parametrization \cite{gm1}, which was adjusted to reproduce nuclear saturation properties employing extra scalar self-meson interactions $\sigma^3$ and $\sigma^4$, together with an additional vector-isovector self-meson $\omega\rho$ interaction, fitted to improve the value reproduced by the symmetry energy at saturation density. The original GM1 set yields an incompressibility modulus of $K=300~\rm MeV$ and a symmetry energy of $S=33~\rm MeV$ at the saturation density. When the
$\omega\rho$ interaction is introduced (called GM1$\omega\rho$ parametrization hereon), 
 the coupling constant $g_{\omega \rho} = 2.015\times 10^{-2}$ is numerically obtained so that the symmetry energy is fixed to be the same as the CMF model described next, i.e., $S=30$, and the remaining parameters are the same as in the standard GM1 parametrization. The GM1 parametrization does not satisfy all the nuclear matter and astrophysical constraints \cite{dutra2014,lourencco2019}, however it is still widely employed and it allows comparisons with a large amount of results available in the literature.
 Recent results indicate that the symmetry energy slope constrained from the neutron skin thickness measurement performed by \cite{prex} can be larger that previously accepted,  reaching values in the range $106 \pm 37$ MeV. If this result is confirmed, the GM1
 parametrization still holds a good prediction power. The inclusion of the crossed interaction between the mesonic fields, as performed here, reduces the symmetry energy slope from $94$ MeV  in the original GM1 parametrization to $69$ MeV in the GM1$\omega\rho$, 
 a value within the usually acceptable range \cite{chen2010,Hebeler:2013nza,dutra2014}. There are other proposals in the literature aiming to reconcile the GM1 model with observables, e.g., in Ref. \cite{malfatti2019hot}  it is suggested a parametrization with the same parameters as GM1, but with a density dependent coupling for the $\rho$ meson field.

Considering $b$ species of baryons, the RMF model is described by the following Lagrangian density 
\begin{eqnarray}
  \mathcal{L}_{\text{RMF}} &=& \sum_{b} \bar{\psi }_{b}\Bigg [ \gamma ^{\mu }\left ( i\partial ^{\mu }-g_{\omega b}\omega _{\mu }-\frac{1}{2}g_{\rho b}\vec{\tau }\cdot \vec{\rho }_{\mu } \right )
  - m_{b}^\ast\Bigg ]\psi _{b}\nonumber \\&+&\frac{1}{2}\left (\partial ^{\mu } \sigma  \partial _{\mu }\sigma - m_{\sigma }^{2}\sigma ^{2} \right ) 
  \nonumber - \frac{\lambda _{1}}{3}\sigma  ^{3}-\frac{\lambda _{2}}{4}\sigma  ^{4} \\&-&\frac{1}{4}\Omega ^{\mu \nu }\Omega _{\mu \nu } +\frac{1}{2}m_{\omega }^{2}\omega _{\mu }\omega ^{\mu }  \nonumber 
-\frac{1}{4}\vec{P}^{\mu \nu }\cdot \vec{P}_{\mu \nu }\\&+&\frac{1}{2}m_{\rho }^{2}\vec{\rho }_{\mu }\cdot \vec{\rho }\,^{\mu } +g_{\omega \rho }\, \omega _{\mu }\omega ^{\mu }\vec{\rho }_{\mu }\cdot\vec{\rho }\,^{\mu } ,
 \label{nlwm}  
\end{eqnarray}
where $m_b^\ast=m_{b}-g_{\sigma b}\sigma$ is the effective mass of the baryon with bare mass $m_b$, $m_i$ is the mass of the meson $i$ with $i=\sigma,\omega,\rho$. The parameter $g_{ib}$ is the coupling constant of the interaction of the $i=\{\sigma$, $\omega$, $\rho\}$ meson field with the baryonic field $\psi_b$. 
Considering the baryons to include nucleons $N=\{p$, $n\}$, hyperons $H=\{\Lambda$, $\Sigma^+$, $\Sigma^0$, $\Sigma^-$, $\Xi^0, \Xi^-\}$ and/or spin $3/2$ resonances $\Delta=\{\Delta^{++}$, $\Delta^{+}$, $\Delta^{0}$, $\Delta^{-}\}$, the interaction coupling between each meson and each hyperon or $\Delta$ can be defined in terms of a scaling of the meson-nucleon coupling $g_{iN}$. In this work, we take the SU(3) hyperon coupling scheme parameters proposed in Ref. \cite{hyperonchi}, and consider the $\Delta$-meson coupling scheme as being given by {{$g_{\omega \Delta}/g_{\omega N}=g_{\rho \Delta}/g_{\rho N}=1$}} and $g_{\sigma \Delta}/g_{\sigma N}=\beta$,  where $\beta$ is a free parameter. The particle relative abundances are very sensitive to both hyperon and delta coupling schemes, but we restrict our analysis to the role of the $\sigma \Delta$ coupling parameter $\beta$, which was shown to have a strong effect on the onset of the $\Delta$ baryons \cite{deltaslo2}.
{For both RMF parameterizations we use, the hyperon and delta potentials are $U_\Lambda=-28$ MeV, $U_\Sigma=32$ MeV, $U_\Xi=39$ MeV, and $U_\Delta =-65$ MeV ($U_\Delta =-93$ MeV) for $\beta=1.0$ ($\beta=1.1$).

\begin{figure*}
\centering
\includegraphics[angle=270,width=8.9cm] {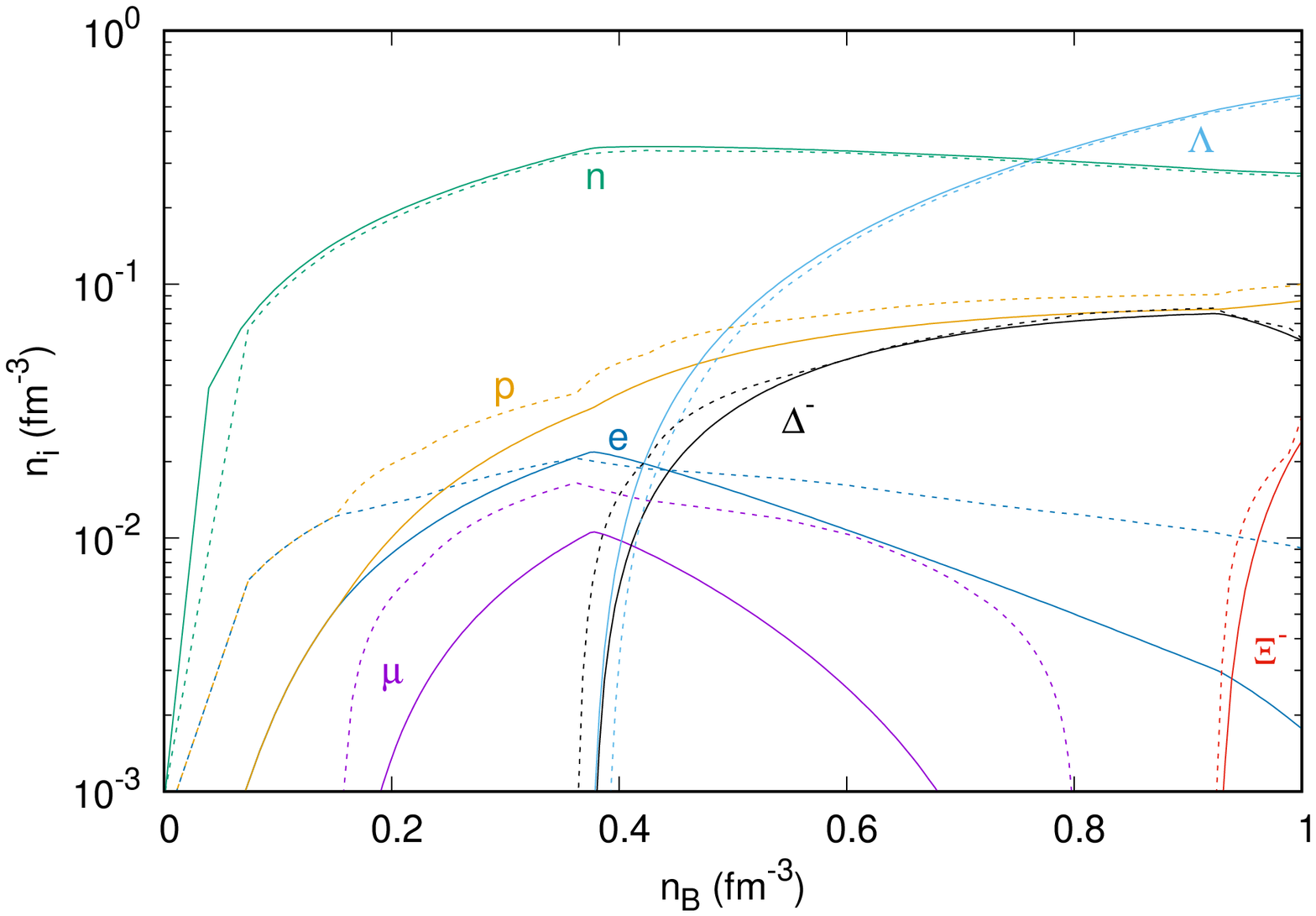}
\includegraphics[angle=270,width=8.9cm] {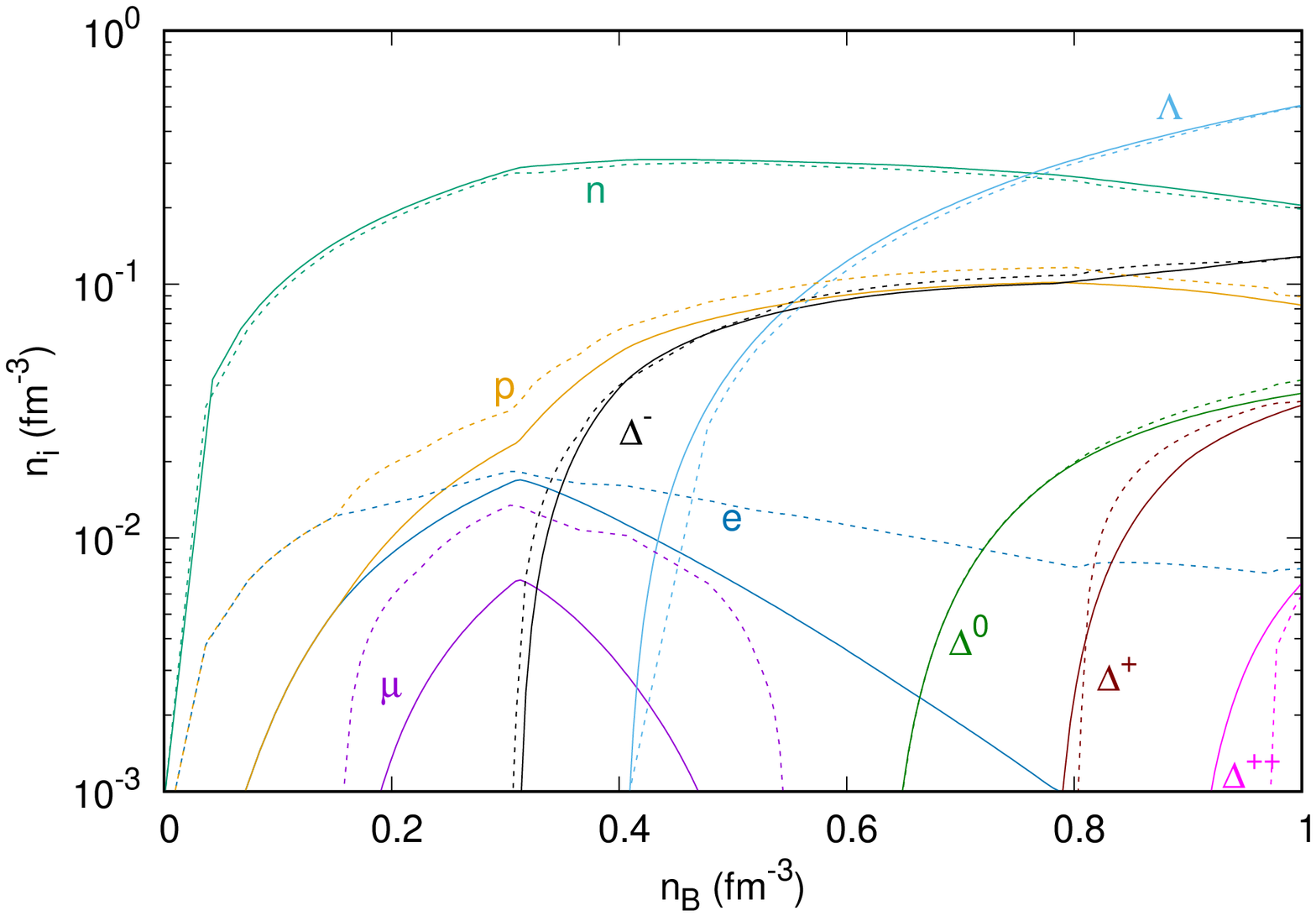}
\caption{The same as in Fig.~1 but using the new GM1$\omega\rho$ parametrization.}
\label{popnlwmwr}
\end{figure*}

The equations of motion for each of the fields are obtained from the Lagrangian, Eq.~(\ref{nlwm}), via the usual Euler-Lagrange formalism, employing a mean-field approximation that allows one to obtain the energy-momentum tensor for this model. 
The energy spectra of the baryons are
\begin{equation}
    E_b^*=E_b-g_{\omega b}\omega_0-\tau_{3b}g_{\rho b}\rho_0,
    \label{ve2}
\end{equation}  
and the Fermi momentum of the charged baryons can be written as
\begin{equation}
    k_z^b=\sqrt{{E_b^*}^2-{m_b^*}^2-2\nu|q_b|B}.
    \label{ve}
\end{equation}
The EoS can be obtained from the energy-momentum tensor, solving numerically the field equations and imposing charge neutrality and chemical equilibrium. In order to fulfill these constraints, a non-interacting lepton gas is included in the description. Then, at zero temperature the particle fractions are determined from the neutron (usually called baryon) and electron chemical potentials:
\begin{eqnarray}
    E_b=\mu_b=\mu_n-q_b\mu_e , \label{betaeq}
\end{eqnarray}
where $q_b$ is the electric charge of the baryon $b$, and $\mu_\mu=\mu_e$. From Eq.~\ref{ve} and using the baryon density or chemical potential as input, we can compute all
$k_z^b$ and the corresponding number of Landau levels \cite{PhysRevC89}.

The effects of the scalar-$\Delta$ coupling are shown in Figs. \ref{popnlwm} and \ref{popnlwmwr}, when comparing the left ($\beta=1.0$) and right ($\beta=1.1$) panels. The increase of this interaction changes the overall $\Delta$-particle threshold to lower densities, while it pushes the hyperon threshold to higher densities. This effect is more pronounced when the $\omega\rho$ interaction and $\beta=1.1$ are used (right panel of Fig. \ref{popnlwmwr}). In this case, the $\Lambda$'s are the only hyperons present, while the four $\Delta$ species appear at relatively low densities. {{ The amount of leptons is reduced significantly.}} For comparison, see Fig.~\ref{popResnlwwm} for a more extreme scenario in which the hyperons were artificially suppressed. In this case, as expected, there is an even larger amount of  different $\Delta$ species at a given density.

\begin{figure*}
\centering
\includegraphics[angle=270,width=8.9cm] {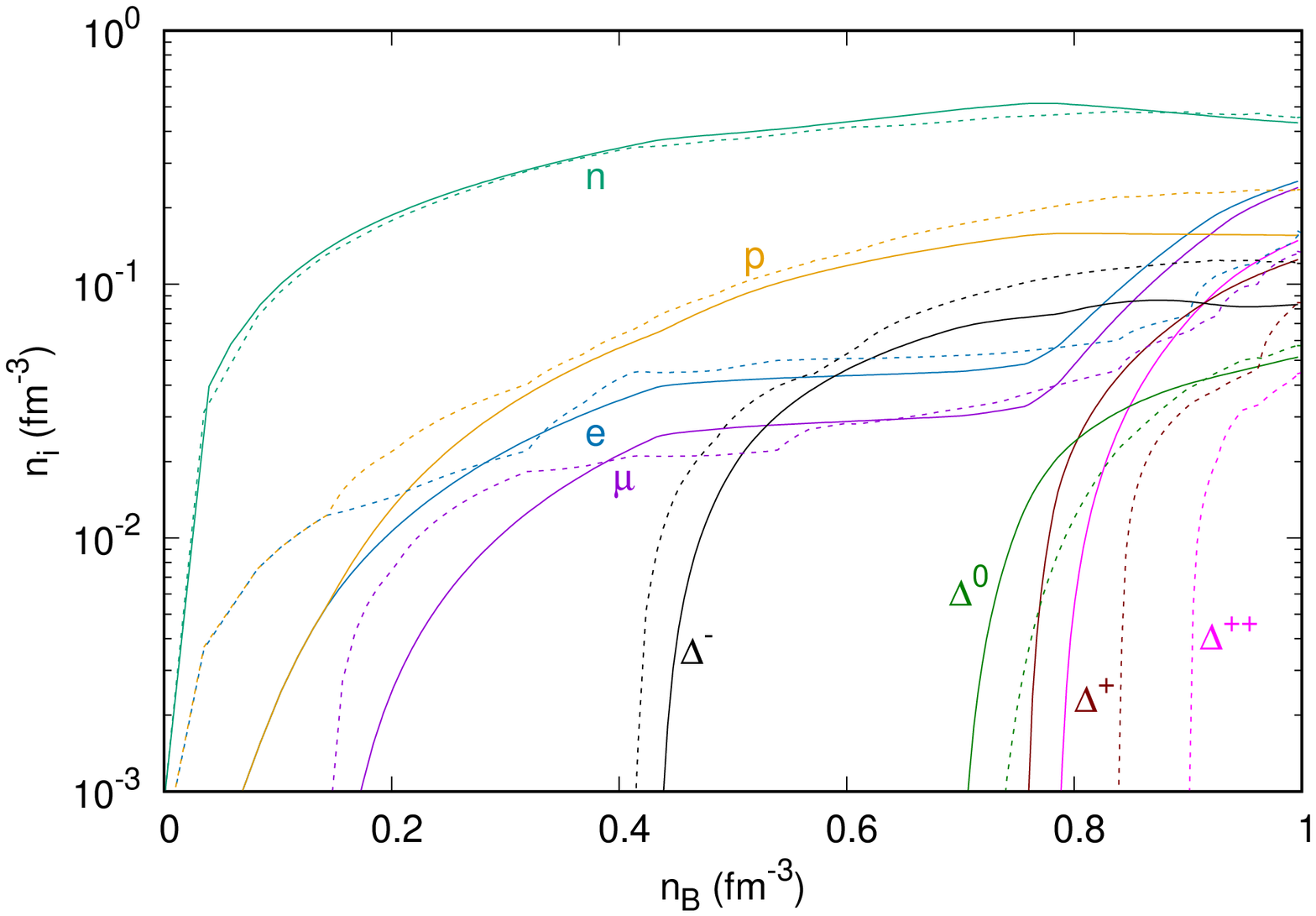}
\includegraphics[angle=270,width=8.9cm] {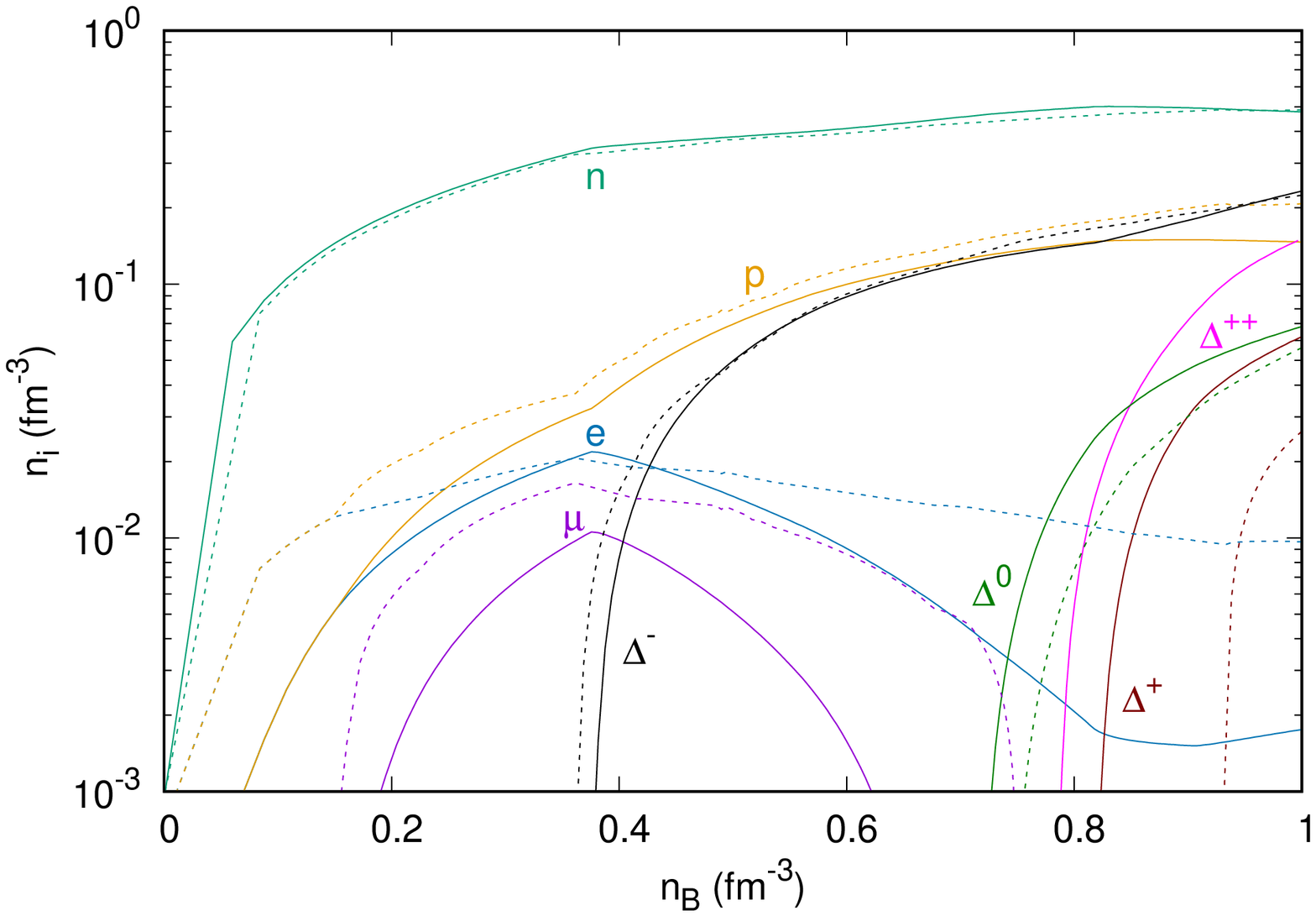}
\caption{Particle population for the RMF model with GM1 parametrization (left panel) and GM1$\omega\rho$ parametrization (right panel), suppressing the hyperons and taking $\beta=1.0$. Once more, full lines show results without magnetic fields, while dashed lines show results including a magnetic field strength of $B=3\times10^{18}$ G.
\label{popResnlwwm}}
\end{figure*}

{{Concerning our choice of values for the parameter $\beta$, they cover the meaningful range that switches from having many hyperons to having many spin $3/2$ baryons. While lower values would completely exclude the $\Delta$'s from our density range, a larger value would completely exclude the hyperons. As seen when comparing both panels of Fig.~1, our choices of $\beta$ are already very close to these extreme cases, in such a way that increasing or decreasing $\beta$ further would not alter significantly our results. On the other hand, in the presence of the $\omega\rho$ coupling, as seen in the left panel of Fig.~2, a lower $\beta$ could decrease further the amount of $\Delta$'s, producing different results.}}

Another interesting feature is the interplay between the $\Xi^-$ and $\Delta^-$ relative populations. In the left panel of Fig. \ref{popnlwm}, one can see that $\Delta^-$'s start to appear at $n_B\approx 0.5$ fm$^{-3}$ (when the magnetic field is not considered), and their population keeps growing with density until it represents  {{$3\%$}} of the total baryon number density $n_B\approx 0.8$ fm$^{-3}$. At this point, the (lighter) $\Xi^-$'s appear and rapidly suppress the other negative particles, dominating over the $\Delta^-$'s soon afterwards. This dynamics is delayed in scenarios where the hyperons are not preferred, e.g., in the scenario shown at the right panel of Fig.~\ref{popnlwmwr}, where the $\Delta$'s show a fast uninterrupted increase with density. This characteristic fast increase occurs for all $\Delta$ baryons when the hyperons are fully suppressed (see Fig.~\ref{popResnlwwm}).

Magnetic field effects on the particle populations are shown using dashed lines in Figs.~1-3. We find that the magnetic field produces a significant enhancement in the amount of most {{charged particles}}, including protons, electrons, muons, $\Sigma^-$'s, and $\Delta^-$'s, but an overall suppression of $\Delta^+$'s and $\Delta^{++}$'s when they are present. The latter is related to the increased amount of protons. In the cases in which hyperons are not present, the magnetic field induces a more significant change in the $\Delta$ population, as shown in Fig.~\ref{popResnlwwm}. The cusps in
the dashed curves correspond to threshold crossings for the maximum Landau levels, known as van Alphen oscillations \cite{oscilation}. In these figures, a constant magnetic field of magnitude $B=3\times 10^{18}$ G was chosen. This corresponds to about the largest magnetic field strength that can be reached in the center of magnetars, as predicted by numerical calculations that solve Einstein's and Maxwell's equations for a pure poloidal configuration \cite{Bocquet:1995je,Cardall:2000bs}. In reality, the magnetic field is not constant within neutron stars, but was found to increase with density by less than one order of magnitude when solving Einstein's and Maxwell's equations for a pure poloidal configuration \cite{Bocquet:1995je,Dexheimer:2016yqu}.

\subsection{Chiral Mean-Field Model}

The chiral mean-field (CMF) model is based on a nonlinear realization of the SU(3) sigma model and is constructed in such a way that chiral symmetry is restored at large temperatures and/or densities. In this work, we restrict ourselves to the hadronic version of the model {{(with leptons)}}, which was fit to reproduce hadronic vacuum masses, decay constants, nuclear saturation properties, and to reach $\sim2.1~\Msun$ stars containing nucleons and hyperons \cite{Dexheimer:2008ax}. The model was also used to investigate the influence of heavier resonances \cite{Schurhoff:2010ph} and magnetic fields \cite{Dexheimer:2011pz,Franzon:2015sya} in neutron stars. More recently, we have modified the original model by including a new vector-isovector self-meson interaction $\omega\rho$ shown to reproduce neutron stars with smaller radius and lower tidal deformability \cite{Dexheimer:2018dhb}. We refit the N$\rho$ coupling constant to reproduce the same symmetry energy as the original model ($S=30$ MeV) and the new $\omega\rho$ coupling constant to reduce significantly the symmetry energy slope and neutron-star radii, but not masses. We have also included a pure-vector self-meson interaction $\omega^4$ that can also be freely varied. This interaction was shown to increase stellar masses \cite{Dexheimer:2020rlp}. See Ref.~\cite{Dexheimer:2020rlp} and references therein for a complete list of coupling constants.
Because in this formalism the interactions are mediated also by the strange $\zeta$ and $\phi$ mesons, the equivalent of Eq.~\ref{ve2} contains an extra ($-g_{\phi b}\phi_0$) term.

\begin{figure*}
\centering
\includegraphics[width=8.9cm,clip,trim=.3cm 0 2.2cm 2.8cm] {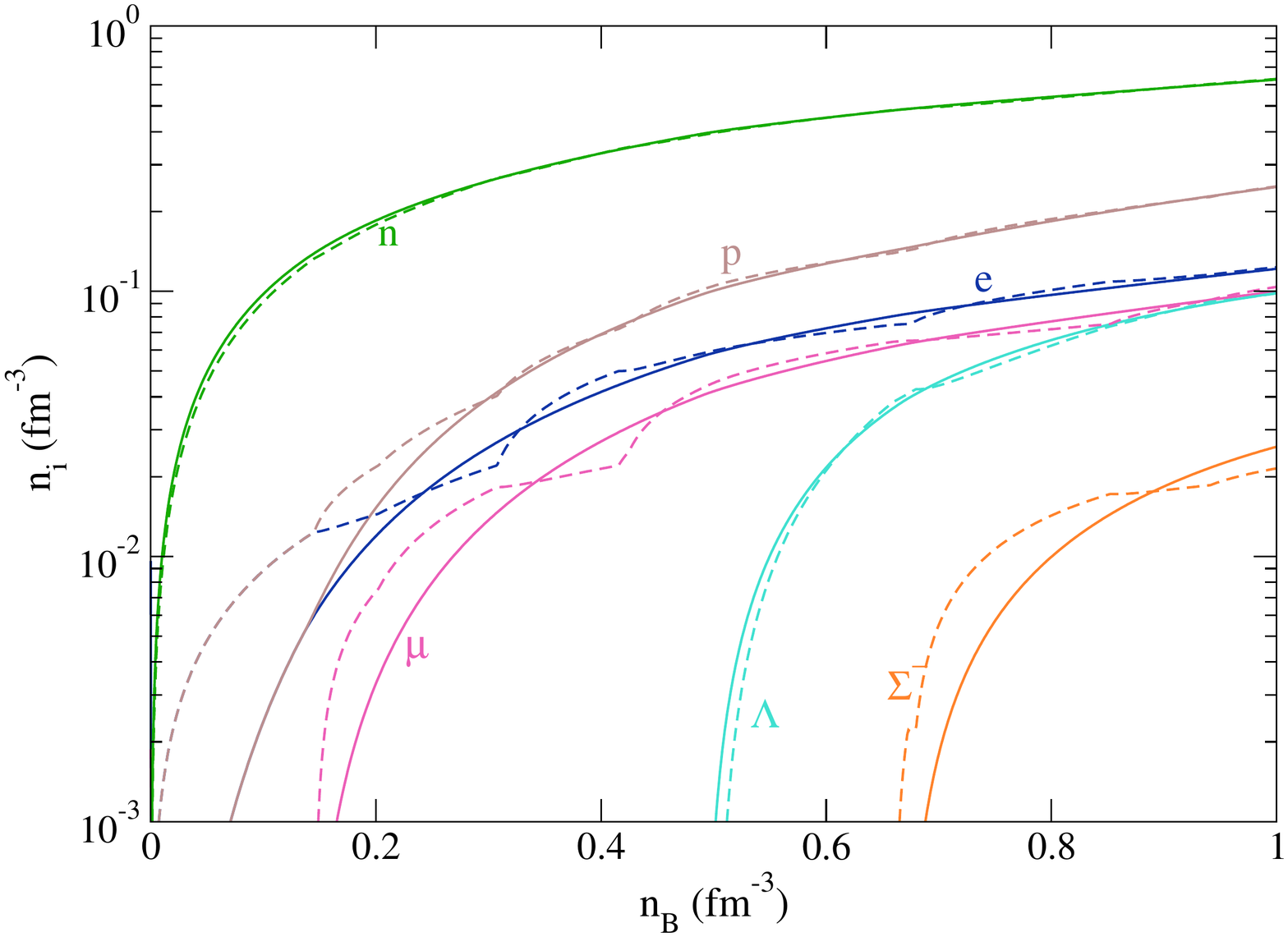}
\includegraphics[width=8.9cm,clip,trim=.3cm 0 2.2cm 2.8cm] {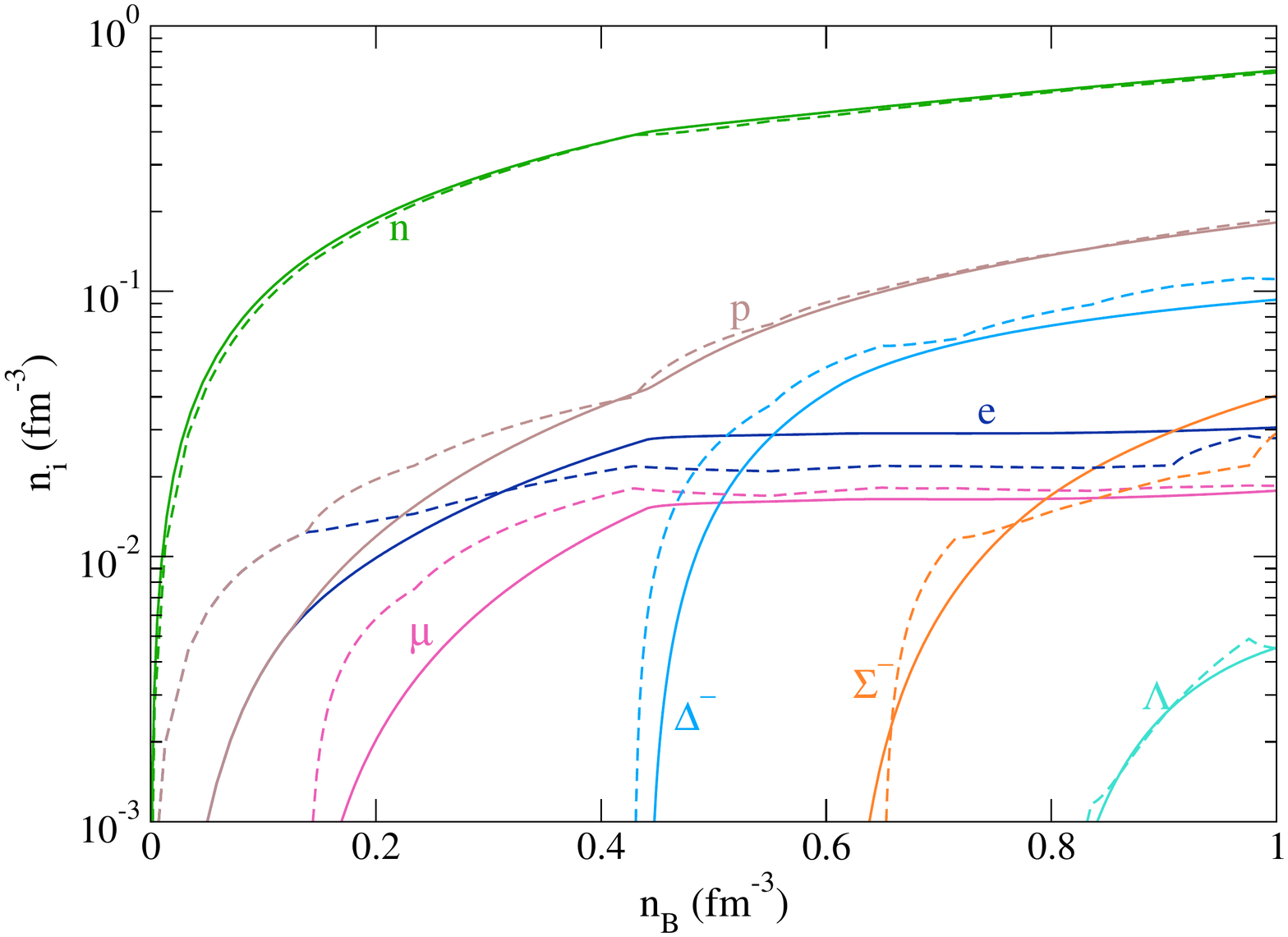}
\caption{Particle population for the CMF model without (left panel) and with (right panel) the $\omega\rho$ interaction as a function of the baryon number density. Full lines show results without magnetic fields, while dashed lines show results including a magnetic field strength of $B=3\times10^{18}$ G.}
\label{popResHyp}
\end{figure*}

\begin{figure*}
\centering
\includegraphics[width=8.9cm,clip,trim=.3cm 0 2.2cm 2.8cm]{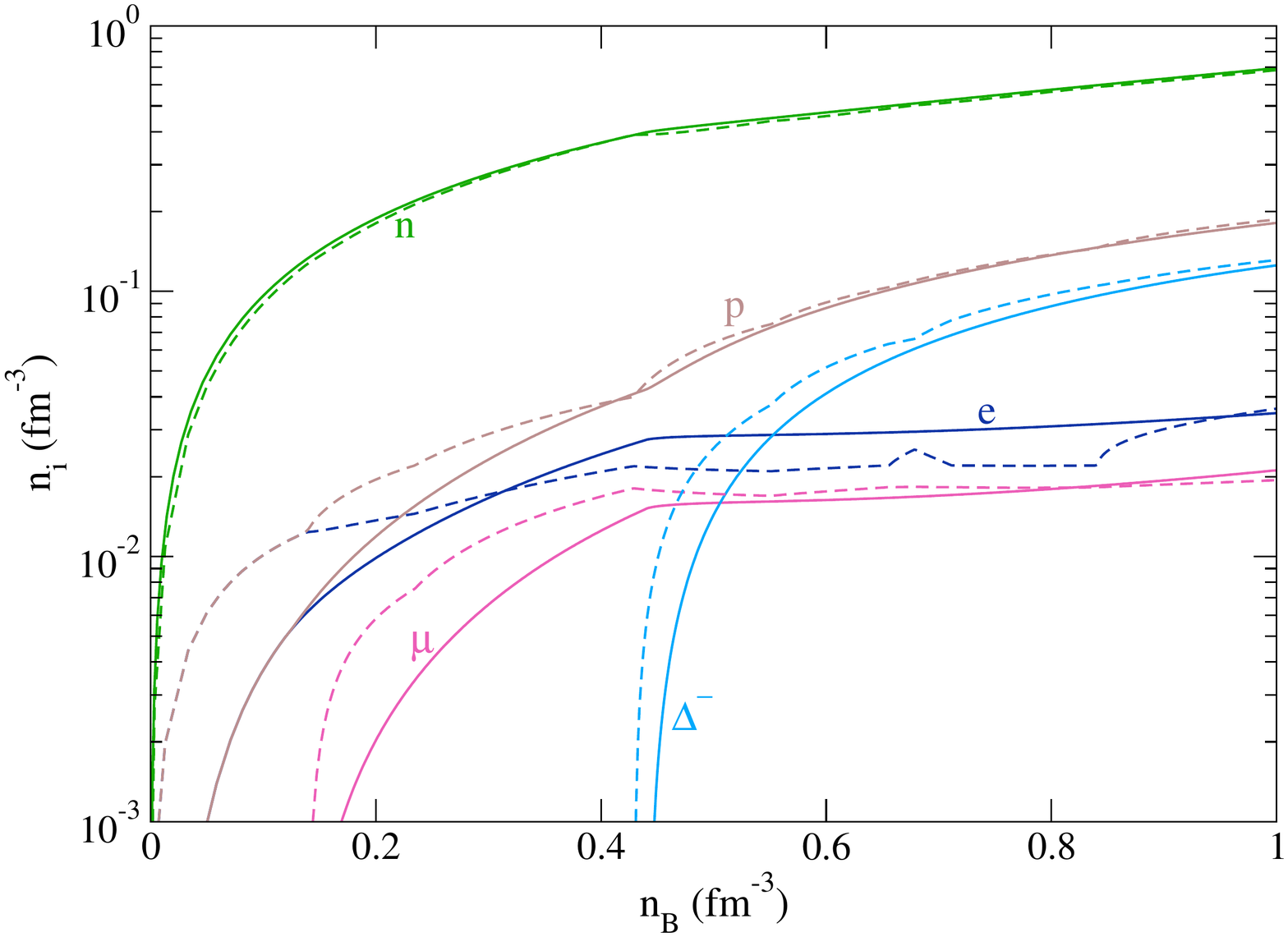}
\includegraphics[width=8.9cm,clip,trim=.3cm 0 2.2cm 2.8cm]{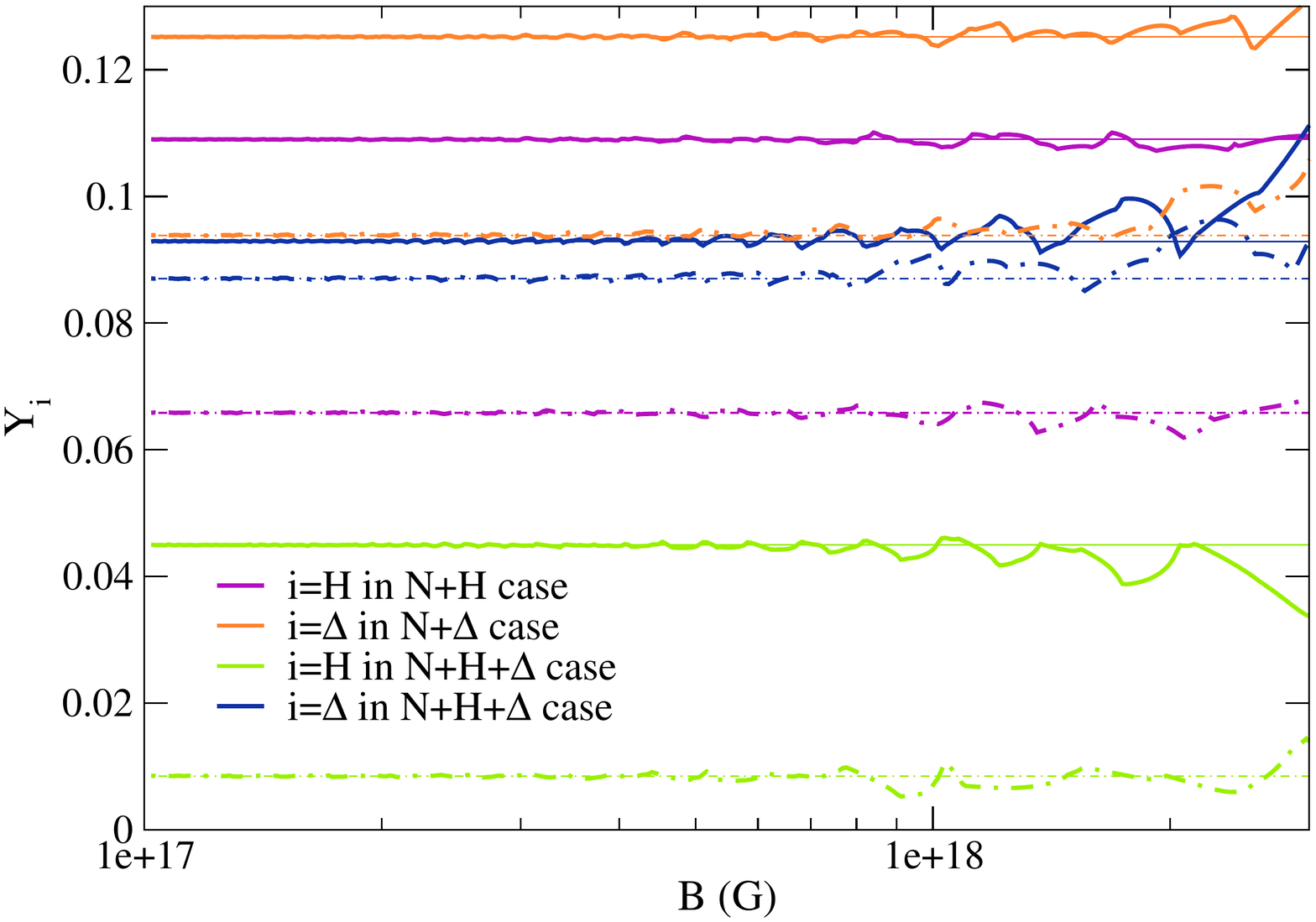}
\caption{Left panel: Same as the right panel of Fig.~\ref{popResHyp} but suppressing the hyperons. Right panel: fraction of hyperons and $\Delta$ baryons as a function of magnetic field strength $B$ at a fixed baryon number densities of $n_B=0.7$ fm$^{-3}$ (dotted-dashed lines) and $n_B=1$ fm$^{-3}$ (full lines), shown for  different configurations. In the right panel, thin lines show zero magnetic field results for comparison.
\label{popRes}}
\end{figure*}

In Fig.~\ref{popResHyp}, we show for the first time the introduction of the $\omega^4$ interaction in the population of pure hadronic matter (with coupling constant $-4.7$), and with additional baryons from the decuplet $\Delta$'s, $\Sigma^*$'s, $\Xi^*$'s, and $\Omega$'s. Note that in Ref.~\cite{Dexheimer:2020rlp} a phase transition to quark matter suppressed most of the hyperons. Here, although we include the whole baryon decuplet, only the $\Delta$'s appear in the relevant regime. Following the SU(3) and SU(6) coupling schemes for the scalar and vector couplings of the mesons to the baryons, there are only two free parameters left: one fitted to reproduce {{a reasonable hyperon potential $U_\Lambda$ and another one ($r_V=g_{N \Delta}/g_{\omega\Delta}=1.25$) chosen to reproduce the potential $U_\Delta\sim U_N$ for symmetric matter at saturation, resulting additionally in $U_\Lambda\sim-27$ MeV, $U_\Sigma=6$ MeV, $U_\Xi=-17$ MeV, and $U_\Delta =-64$ MeV (in the presence of the  additional interaction $\omega^4$).}}
A much larger $r_V$ would suppress all $\Delta$'s, while a much lower value would suppress all hyperons.

The difference between the two panels in Fig.~4 is only due to the addition of the $\omega\rho$ interaction (with normalized coupling constant $0$ on the left panel and $62$ in the right panel). This interaction generates matter with a more soft symmetry energy beyond saturation (lower value for slope), meaning a lower energy cost to produce isospin and, therefore, a larger neutron-to-proton ratio, more $\Sigma^-$'s and $\Delta^-$'s, but less leptons and $\Lambda$'s. As a result, the $\Delta^-$'s appear before any hyperon species. The very slow increase of $\Sigma^-$'s with density (for zero magnetic field) barely affects the $\Delta^-$ population.

\begin{figure*}
\centering
\includegraphics[angle=270,width=8.9cm]{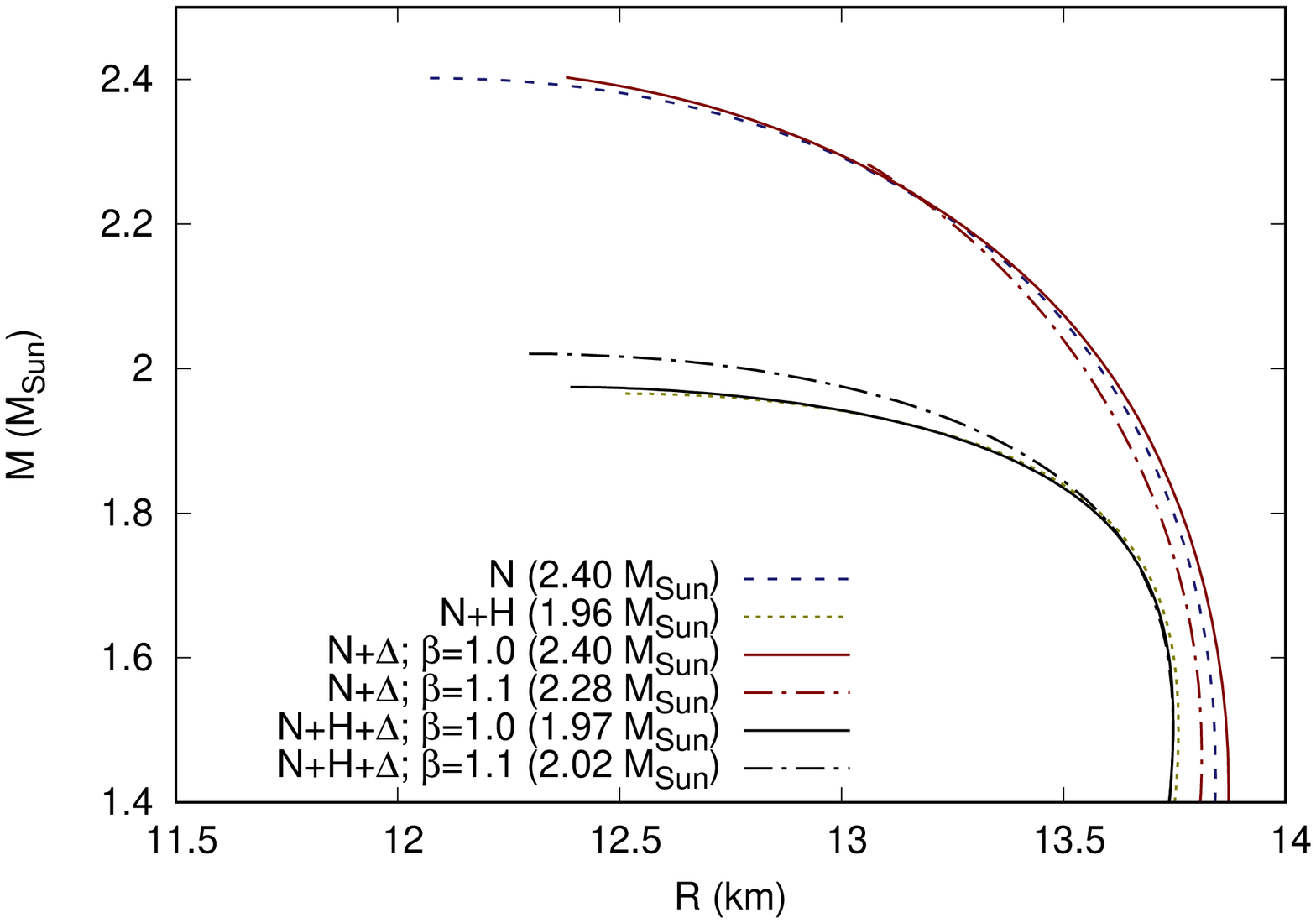}
\includegraphics[angle=270,width=8.9cm]{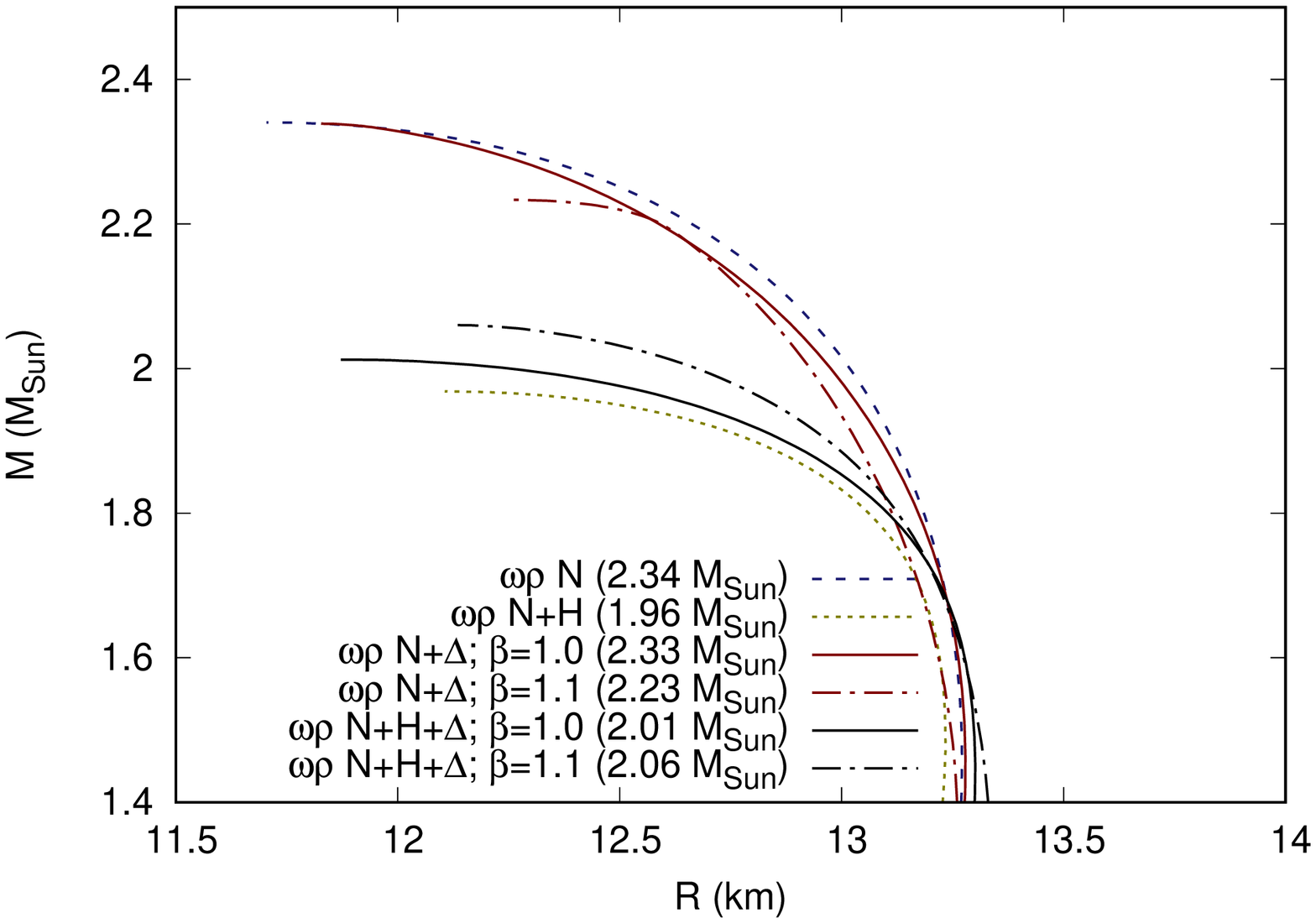}
\caption{Mass-radius diagram for the RMF model with GM1 parametrization (left panel) and  GM1$\omega\rho$ parametrization (right panel), showing results without magnetic field effects. The maximum stellar masses are indicated for all cases.}
\label{TOVnlwm}
\end{figure*}

Continuing our discussion with only configurations that include the $\omega\rho$ interaction, the left panel of Fig.~\ref{popRes} illustrates the case in which hyperons are suppressed. Note that not even in this case the other baryons from the decuplet appear in the density regime relevant for neutron stars, although a larger amount of $\Delta^-$'s appear at large densities. As shown by the dashed lines of Fig.~4 and left panel of Fig.~5, the magnetic field further enhances the amount of negatively charged $\Delta$'s. The right panel of Fig.~\ref{popRes} illustrates how the magnetic field changes the relative density (fraction) of hyperons

\begin{eqnarray}
{{Y_H=\frac{n_{\Lambda} +n_{\Sigma^+}+n_{\Sigma^0}+n_{\Sigma^-}+n_{\Xi^0}+n_{\Xi^-}}{n_B} ,}}\nonumber \\
\end{eqnarray}
and fraction of $\Delta$'s
\begin{eqnarray}
{{Y_\Delta=\frac{ n_{\Delta^{++}}+n_{\Delta^+}+n_{\Delta^0}+n_{\Delta^-}}{n_B} ,}}
\end{eqnarray}
at different large densities. The thin baselines show the zero magnetic field value for comparison. In this panel it can be seen that the magnetic field enhances the fraction of $\Delta$'s (orange and navy blue lines), while suppressing the total fraction of hyperons (violet and green lines) for all the cases studied. {{The Van Alphen oscillations can be very easily identified for large values of magnetic field.}}

The CMF model EoS's presented in this work, together some other variations, have been added to the CompOSE online repository\footnote{http://compose.obspm.fr} under the category ``Cold Neutron Star EoS". Microscopic quantities used for transport calculations were also provided. To our knowledge, these are the only EoS's containing $\Delta$ baryons that appear in the CompOSE repository at the moment.

\section{Results: Neutron Stars}

In this section, we discuss the results of inserting our EoS's and some additional ones (with only nucleons) in the Tolman–Oppenheimer–Volkoff equations \cite{Tolman:1939jz,Oppenheimer:1939ne}. For the crust, we use the BPS EoS \cite{bps}. We only analyze mass-radius curves produced from EoS's without magnetic field effects, with the purpose of comparing different cases and parametrizations with observational data. We leave the study of mass-radius relations for magnetic neutron stars to a further publication, as this requires the solution of a more complicated system of equations in general relativity \cite{Bocquet:1995je,Cardall:2000bs,Frieben:2012dz,Pili:2014npa}.

In Fig. \ref{TOVnlwm}, we show RMF model results without (left panel) and with (right panel) the $\omega\rho$ interaction. In both cases, the pure nucleonic stars are more massive, and the ones that contain hyperons less massive. But, more interestingly,  stars that include all degrees of freedom are more massive than the ones that include only nucleons and hyperons, and this effect is more obvious for the case with $\omega\rho$ interaction and $\beta=1.1$. When the hyperons are suppressed, the stellar masses increase even more, in some cases surpassing the nucleonic star masses.  

This discussion is related to the well-known hyperon puzzle \cite{Chamel:2013efa}, but with a twist. As shown in Fig. \ref{sfnlwm}, going from dashed lines to solid or dot-dashed, the addition of $\Delta$'s decreases the fraction of nucleons (some neutrons) and hyperons ($\Lambda$'s) to create $\Delta$'s and some protons. The overall increase in isospin asymmetry makes the EoS stiffer, even when more degrees of freedom are present. The larger population change caused by the $\omega\rho$, $\beta=1.1$ parametrization only enhances this effect. See Ref.~\cite{Li:2020ias} for a detailed study of the effect of {{the}} symmetry energy at larger densities on the properties of neutron stars with $\Delta$ baryons. Concerning stellar radius modifications, the intermediate density EoS softening caused by the $\omega\rho$ interaction turns stars smaller in all cases.  
Lower radii for $\sim1.4~\Msun$ stars improves the agreement of this model with NICER and LIGO data \cite{Miller:2019cac,Riley:2019yda,LIGOGW170817}.


\begin{figure}
\centering
\includegraphics[angle=270,width=8.9cm]{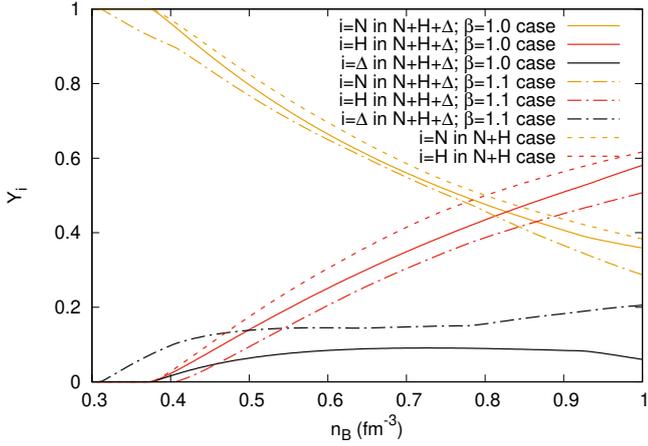}
\caption{Fraction of nucleons, hyperons and $\Delta$'s for the RMF model with GM1$\omega\rho$ parametrization when considering a N+H (dashed lines) or a N+H+$\Delta$ population, with $\beta=1.0$ (solid lines) or $\beta=1.1$ (dashed-dotted lines). Magnetic field effects are not included.}
\label{sfnlwm}
\end{figure}

\begin{figure}
\begin{center}
\includegraphics[width=8.5cm,clip,trim=.3cm 0 2.2cm 2.5cm]{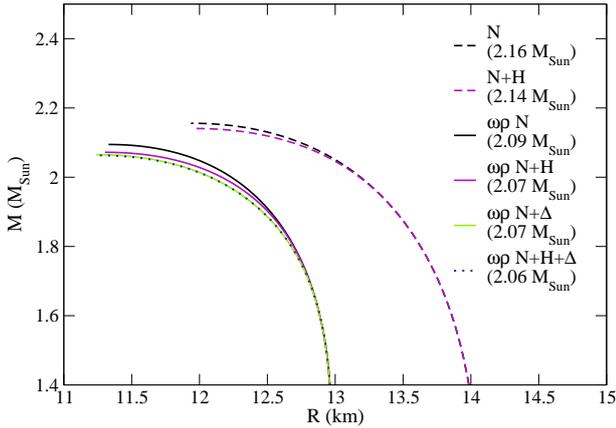}
\caption{Mass-radius diagram for the CMF model showing results without magnetic field effects. The maximum stellar masses are indicated for all cases.}
\label{TOV}
\end{center}
\end{figure}

In Fig. \ref{TOV}, the dashed curves show results without the $\omega\rho$ interaction for the CMF model. In this case, for our selected parametrization, the $\Delta$'s are never present. We show results for nucleons only, and also including hyperons, when the maximum mass decreases. The remaining curves show results with the $\omega\rho$ interaction. In this case, the radii of the $\sim1.4~\Msun$ stars and respective tidal deformabilities are again in much better agreement with NICER and LIGO data. The pure nucleonic stars are more massive, followed by the stars also including hyperons, also including $\Delta$'s and, finally, including all degrees of freedom, when the maximum allowed mass decreases still by only a small amount. The difference when compared to the RMF model discussion in the previous paragraph is that not many hyperons appear in the CMF model in any case, so the suppression (mainly of $\Sigma^-$'s) caused by the appearance of $\Delta$'s is small, and does not change significantly the isospin amount. From the right panel of Fig.~5 it can be seen that the overall number of exotic baryons i) H's+$\Delta$'s in the N+H+$\Delta$ case, then ii) $\Delta$'s in the N+$\Delta$ case, then the iii) H's in the N+H case, correspond inversely to the maximum masses of neutron stars for each case. This dynamics works in the same ways as the hyperon puzzle, but extended to include $\Delta$'s.

\section{Conclusions}

In this work we used two relativistic mean field models to study the influence of magnetic fields in dense matter containing hyperons and $\Delta$ baryons. We also analyzed the effect of including a new vector-isovector interaction on the particle population. We found that in both models this interaction enhances the amount of $\Delta$'s, and at the same time reproduces smaller, but still massive neutron stars. The presence of a strong magnetic field only enhances further the amount of $\Delta$'s is dense matter, but we leave the more complex analysis of macroscopic {{stellar}} properties to a future publication. 

We have {{also not focused on the discussion of the equation of state stiffness in the presence of a strong magnetic field}}. This is because changes in pressure in different directions (with respect to the magnetic field) caused by matter is sub-leading in comparison to changes caused by the Lorentz force \cite{Chatterjee:2014qsa,Franzon:2015sya}. The solution of Einstein's and Maxwell's equations shows that the maximum stable neutron star masses increase as a function of the magnetic field, but the radius change depends on the strength of the poloidal versus the toroidal components \cite{Pili:2014npa}. The magnetic field also influences the calculation of the tidal deformability \cite{Biswas:2019gkw}.

Note that in Ref.~\cite{Thapa:2020ohp}, it is stated that the magnetic field causes hyperons to appear earlier, while the $\Delta$'s are suppressed by the magnetic field, which is opposite to what we found. In this work, we have not included anomalous magnetic moment corrections. Although they were shown to have non-negligible corrections to magnetic field effects in neutron stars \cite{Broderick:2001qw}, in this work we focused on $\Delta$ baryon, and there is very little data for the anomalous magnetic moment coupling strength of these particles \cite{Zyla:2020zbs}. 

\section*{Acknowledgements}

We thank Madappa Prakash for very useful discussions. Support for this research comes from the National Science Foundation under grant PHY-1748621 and PHAROS (COST Action CA16214). This  work is a part of the project INCT-FNA Proc. No. 464898/2014-5. D.P.M. and K.D.M. are  partially supported by Conselho Nacional de Desenvolvimento Cient\'ifico  e  Tecnol\'ogico  (CNPq/Brazil)  respectively  under grant  301155.2017-8  and  with  a  doctorate  scholarship.

\bibliographystyle{unsrt}
\bibliography{paper}

\end{document}